\documentstyle[twoside,fleqn,espcrc2,epsfig]{article}

\newcommand{\AmS}{{\protect\the\textfont2
  A\kern-.1667em\lower.5ex\hbox{M}\kern-.125emS}}

\newcommand{\nub}{\overline{\nu}}

\def\nim#1#2#3  {{\it Nucl. Instr. Meth.} {\bf#1}, #2 (#3). }
\def\np#1#2#3   {{\it Nucl. Phys.} {\bf#1}, #2 (#3). }
\def\pcps#1#2#3 {{\it Proc. Cam. Phil. Soc.} {\bf#1}, #2 (#3). }
\def\pl#1#2#3   {{\it Phys. Lett.} {\bf#1}, #2 (#3). }
\def\plc#1#2#3   {{\it Phys. Lett.} {\bf#1}, #2 (#3); }
\def\prep#1#2#3 {{\it Phys. Rep.} {\bf#1}, #2 (#3). }
\def\prev#1#2#3 {{\it Phys. Rev.} {\bf#1}, #2 (#3). }
\def\prl#1#2#3  {{\it Phys. Rev. Lett.} {\bf#1}, #2 (#3). }
\def\prs#1#2#3  {{\it Proc. Roy. Soc.} {\bf#1}, #2 (#3). }
\def\rmp#1#2#3  {{\it Rev. Mod. Phys.} {\bf#1}, #2 (#3). }
\def\rpp#1#2#3  {{\it Rep. Prog. Phys.} {\bf#1}, #2 (#3). }
\def\zp#1#2#3   {{\it Z. Phys.} {\bf#1}, #2 (#3). }
\def\epj#1#2#3   {{\it Eur. Phys. Jour.} {\bf#1}, #2 (#3). }

\begin{document}
\hyphenation{author another created financial paper re-commend-ed}

\title {Recent structure function results from neutrino scattering at Fermilab
\thanks{To be published 
in proceedings of the XXXth INternational Conferencee on High Energy Physics,
Osaka, Japan, July, 2000}}

\author{U.~K.~Yang,$^{7}$ T.~Adams,$^{4}$ A.~Alton,$^{4}$
C.~G.~Arroyo,$^{2}$ S.~Avvakumov,$^{7}$ L.~de~Barbaro,$^{5}$
P.~de~Barbaro,$^{7}$ A.~O.~Bazarko,$^{2}$ R.~H.~Bernstein,$^{3}$
A.~Bodek,$^{7}$ T.~Bolton,$^{4}$ J.~Brau,$^{6}$ D.~Buchholz,$^{5}$
H.~Budd,$^{7}$ L.~Bugel,$^{3}$ J.~Conrad,$^{2}$ R.~B.~Drucker,$^{6}$
B.~T.~Fleming,$^{2}$
J.~A.~Formaggio,$^{2}$ R.~Frey,$^{6}$ J.~Goldman,$^{4}$
M.~Goncharov,$^{4}$ D.~A.~Harris,$ ^{7} $ R.~A.~Johnson,$^{1}$
J.~H.~Kim,$^{2}$ B.~J.~King,$^{2}$ T.~Kinnel,$^{8}$
S.~Koutsoliotas,$^{2}$ M.~J.~Lamm,$^{3}$ W.~Marsh,$^{3}$
D.~Mason,$^{6}$ K.~S.~McFarland, $^{7}$ C.~McNulty,$^{2}$
S.~R.~Mishra,$^{2}$ D.~Naples,$^{4}$  P.~Nienaber,$^{3}$
A.~Romosan,$^{2}$ W.~K.~Sakumoto,$^{7}$ H.~Schellman,$^{5}$
F.~J.~Sciulli,$^{2}$ W.~G.~Seligman,$^{2}$ M.~H.~Shaevitz,$^{2}$
W.~H.~Smith,$^{8}$ P.~Spentzouris, $^{2}$ E.~G.~Stern,$^{2}$
N.~Suwonjandee,$^{1}$ A.~Vaitaitis,$^{2}$ M.~Vakili,$^{1}$   
J.~Yu,$^{3}$ G.~P.~Zeller,$^{5}$ and E.~D.~Zimmerman$^{2}$ \\
~~~~~~~~~~~~~~~~~~~~~~
(Presented by Un-ki Yang for the CCFR/NuTeV Collaboration)\\
$^{1}$ University of Cincinnati, Cincinnati, OH 45221 \\
$^{2}$ Columbia University, New York, NY 10027 \\
$^{3}$ Fermi National Accelerator Laboratory, Batavia, IL 60510 \\
$^{4}$ Kansas State University, Manhattan, KS 66506 \\
$^{5}$ Northwestern University, Evanston, IL 60208 \\
$^{6}$ University of Oregon, Eugene, OR 97403 \\
$^{7}$ University of Rochester, Rochester, NY 14627 \\
$^{8}$ University of Wisconsin, Madison, WI 53706 }

\begin{abstract}
We report on the extraction of  the structure functions
$F_2$ and  $\Delta xF_3 = xF_3^{\nu}-xF_3^{\nub}$
from CCFR  $\nu_\mu$-Fe and $\nub_\mu$-Fe differential cross sections.
The extraction is performed in a physics model independent (PMI) way.
This first measurement of  $\Delta xF_3$, which is useful in testing
models of heavy charm production, is higher
than current theoretical predictions.
The ratio of the $F_2$ (PMI) values 
measured in $\nu_\mu$ and $\mu$ scattering
is in agreement  (within 5\%) with the NLO predictions 
using massive charm production schemes, thus
resolving the long-standing discrepancy between the two
sets of data. In addition, measurements of $F_L$ (or, equivalently, $R$)
and $2xF_1$ are reported
in the kinematic region where  anomalous nuclear effects in $R$ are 
observed at HERMES.
[Preprint UR-1614, ER/40685/952]
\end{abstract}

\maketitle

Deep inelastic lepton-nucleon scattering experiments have been used
to determine the quark distributions in the nucleon.
However, the quark distributions determined from $\mu$
and $\nu$
experiments\cite{NMC,SEL} were found to be different at small values of $x$,
because of a disagreement in the extracted structure functions.
Here,
we find that the neutrino-muon difference is resolved by
extracting the $\nu_\mu$ structure functions from CCFR neutrino data
in a physics model independent (PMI) way. In addition,
measurements of $\Delta xF_3$, $F_L$, and $2xF_1$ are presented.

The sum of $\nu_\mu$ and $\nub_\mu$
 differential cross sections 
for charged current interactions on an isoscalar target is related to the
structure functions as follows:
\begin{tabbing}
$F(\epsilon)$ \= $\equiv \left[\frac{d^2\sigma^{\nu }}{dxdy}+
\frac{d^2\sigma^{\overline \nu}}{dxdy} \right] 
 \frac {(1-\epsilon)\pi}{y^2G_F^2ME_\nu}$ \\
  \> $ = 2xF_1 [ 1+\epsilon R ] + \frac {y(1-y/2)}{1+(1-y)^2} \Delta xF_3. $~~~  (1)
\end{tabbing}
Here $G_{F}$ is the Fermi weak coupling constant, $M$ is the nucleon
mass, $E_{\nu}$ is the incident energy, the scaling
variable $y=E_h/E_\nu$ is the fractional energy transferred to
the hadronic vertex, $E_h$ is the final state hadronic
energy, and $\epsilon\simeq2(1-y)/(1+(1-y)^2)$ is the polarization of 
the virtual $W$ boson.
The structure
function $2xF_1$ is expressed in terms of $F_2$
by $2xF_1(x,Q^2)=F_2(x,Q^2)\times
\frac{1+4M^2x^2/Q^2}{1+R(x,Q^2)}$, where $Q^2$ is the
square of the four-momentum transfer to the nucleon,
  $x=Q^2/2ME_h$ is
the fractional momentum carried by the struck quark,
and $R=\frac{\sigma
_{L}}{\sigma _{T}} $ is the ratio of the cross-sections of
longitudinally- to transversely-polarized $W$ bosons.
The $\Delta xF_3$ term, which in leading order
 $\simeq 4x(s-c)$, is not present in the $\mu$-scattering case.
In addition, there is a threshold suppression originating
from the production of heavy $c$ quarks 
in a  $\nu_\mu$ charged current interaction with
$s$ quarks. For $\mu$-scattering, there is no suppression for scattering
from $s$ quarks, but more suppression when scattering
from $c$ quarks.

In previous analyses
of $\nu_\mu$ data\cite{SEL},
structure functions were extracted
by applying a slow rescaling correction to correct for the charm
mass suppression in the final state. In addition, 
the $\Delta xF_3$ term from a leading order charm production model
was used as input in the extraction.
These resulted in physics model dependent (PMD) structure functions\cite{SEL}. 
In the new analysis reported here, slow rescaling corrections are not applied.
$\Delta xF_3$ and $F_2$
are extracted from two-parameter fits to the $F(\epsilon)$ distributions
according to Eq. (1).
However, in the $x>0.1$ region,
we extract values of $F_2$ with $\Delta xF_3$ constrained 
to the NLO TR-VFS(MRST)\cite{tr-vfs} predictions. Since 
 $\Delta xF_3$ for $x>0.1$ is small, the extracted
values of $F_2$ are insensitive to $\Delta xF_3$.

\begin{figure}[t]
\centerline{\psfig{figure=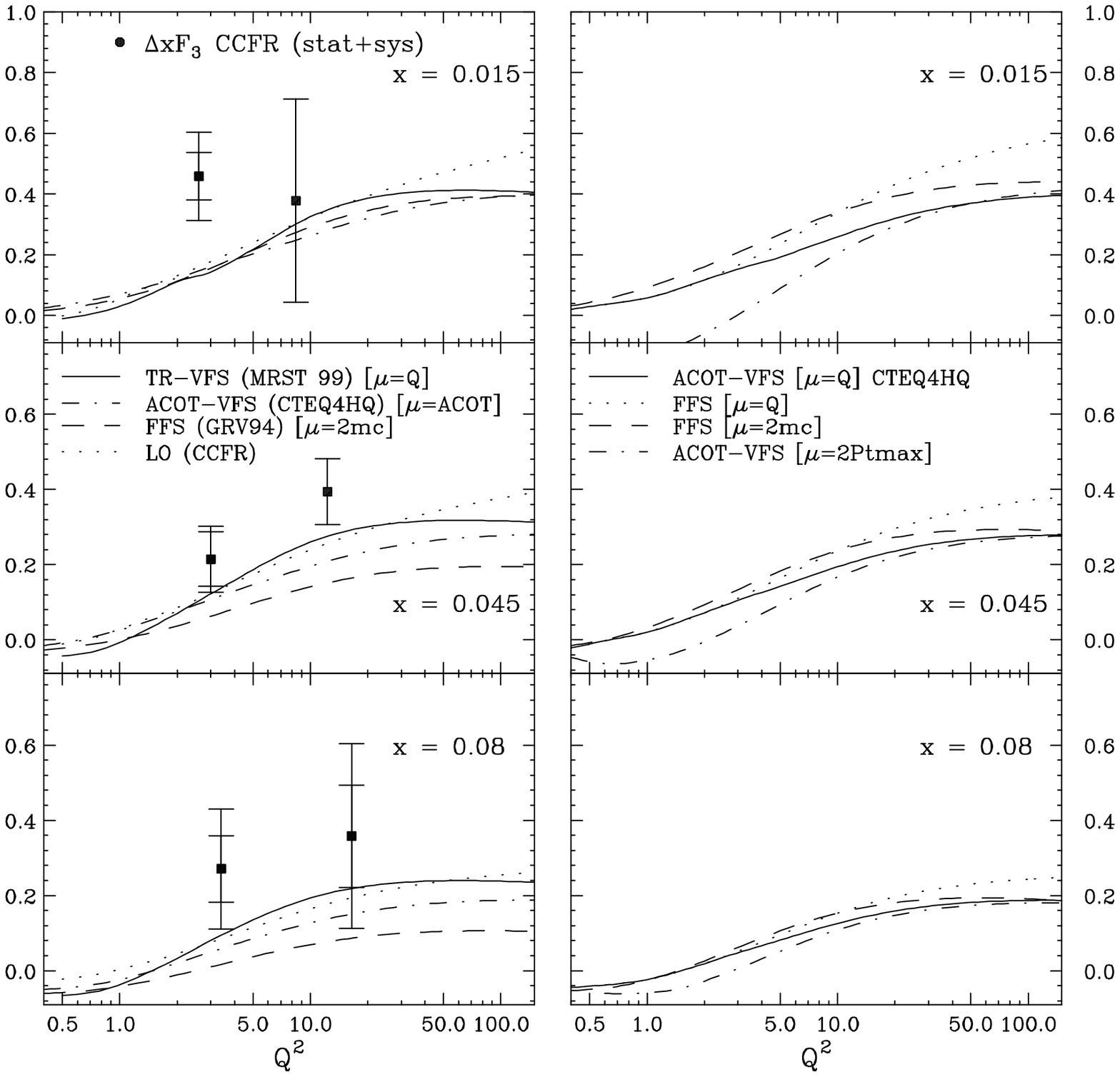,width=3.0in}}
\caption{ $\Delta xF_3$ data as a function of $x$ (above $Q^2=1$) 
compared with various schemes for massive charm production. 
(Left) TR-VFS(MRST99), ACOT-VFS(CTEQ4HQ), FFS(GRV94), and the CCFR-LO (a leading order model with a slow rescaling correction):
(right) sensitivity of the theoretical calculations
to the choice of scale.}
\label{fig:dxf3}
\end{figure}

Fig.~\ref{fig:dxf3}(left)   shows the extracted
values of $\Delta xF_3$ as a function of $x$  (above $Q^2=1$ ), including both statistical
and systematic errors, compared to
various theoretical methods for modeling
heavy charm productions within a QCD framework. 
Fig.~\ref{fig:dxf3}(right) shows the sensitivity to
the choice of scale.
With reasonable choices of scale, all the theoretical models
yield similar results. However,
at low $Q^2$,  our $\Delta xF_3$ data are higher than all
theoretical models.

\begin{figure}[t]
\centerline{\psfig{figure=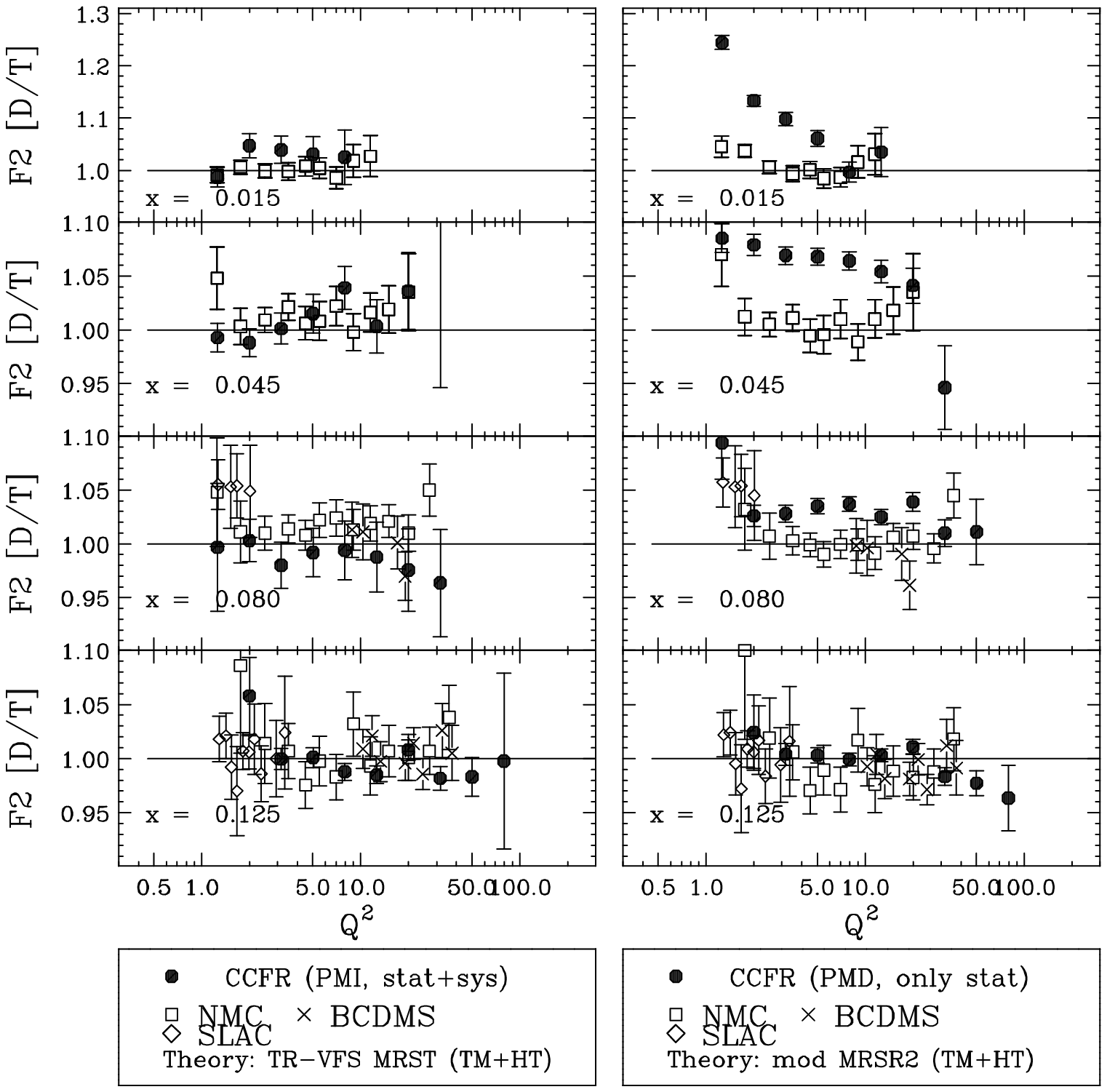 ,width=3.0in}}
\caption{ (Left) The ratio of the
$F_2^{\nu}$ (PMI) data divided by the predictions of TR-VFS (MRST99)
with target mass and higher twist corrections;
(right) The ratio of the previous
$F_2^{\nu}$ (PMD) data and the predictions of MRSR2. 
Also shown are the ratios of
the $F_2^{\mu}$ (NMC, BCDMS) and $F_2^e$ (SLAC) to the theoretical predictions.}
\label{fig:f2}
\end{figure}

Our $F_2$ (PMI)
measurements divided by 
the NLO TR-VFS(MRST) predictions are shown in Fig.~\ref{fig:f2}(left). Also shown are $F_2^{\mu}$ and
$F_2^e$   divided
by the theory predictions. Nuclear effects, target mass, and
higher twist corrections are included in the calculation.
As shown in Fig.~\ref{fig:f2}, within 5\%
both the neutrino and muon
structure functions are in agreement with the NLO TR-VFS(MRST) predictions,
and therefore in agreement with each other, thus
resolving the long-standing discrepancy between the two
sets of data. A comparison
using the NLO ACOT-VFS(CTEQ4HQ)\cite{vfs} predictions yields similar results.
Note that previously there was up to a 20\% difference 
between the CCFR $F_2$ (PMD) 
and NMC data at  $x=0.015$, as shown in Fig.~\ref{fig:f2}(right).

 Recently, there has been a renewed
interest in $R$ at small $x$ and $Q^2<1$, because of the large anomalous nuclear 
effect that has been reported by the
HERMES experiment\cite{hermes}. Their measurement implies a large enhancement in $F_L$
but suppression in $2xF_1$ in heavy nuclear targets.
It is expected that any nuclear effect
in $R$ would be enhanced in the CCFR iron target
with respect to the nitrogen target in HERMES,  unless the origin of this effect
depends on the incident probe (electron versus neutrino).

Values of $F_L$
and $2xF_1$ are extracted from the sums of
the corrected $\nu_\mu$-Fe and $\nub_\mu$-Fe
differential cross sections in different
energy bins according to Eq.~(1).
An extraction of $F_L$ requires knowledge of $\Delta xF_3$.
which we obtain from  the NLO TR-VFS(MRST) calculation.
Because of the large uncertainty in $\Delta xF_3$ at low $Q^2$ region,
an extrapolation of the curve which describes the measured CCFR $\Delta xF_3$ data above $Q^2=1$
is used for the systematic error. Here we are interested in the relative $Q^2$ dependence
of $F_L$ and $2xF_1$.

\begin{figure}[t]
\centerline{\psfig{figure=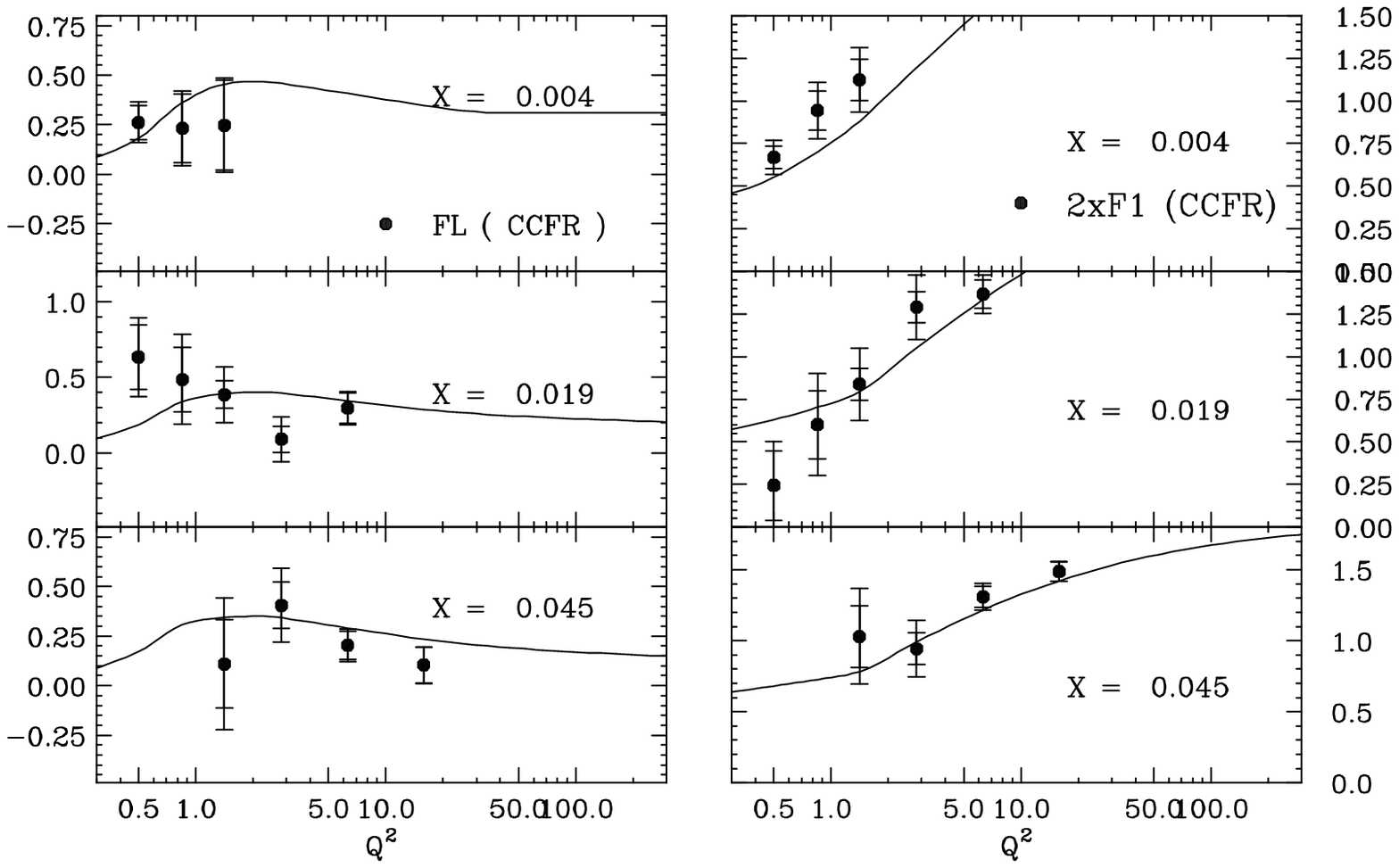,width=2.8in}}
\caption{Preliminary measurements of $F_L$ and $2xF_1$
as a function of $Q^2$ for $x<0.1$, The curves are the predictions from a QCD inspired leading
order fit to the CCFR differential cross section
data with $R$= $R_{world}$ for neutrino scattering.}
\label{fig:FL_2xF1}
\end{figure}

Fig.~\ref{fig:FL_2xF1} shows the preliminary values of $F_L$ and $2xF_1$
as a function of $Q^2$ for $x<0.1$. The inner errors include 
both statistical and experimental
systematic errors. The outer errors represent
the $\Delta xF_3$ model errors added in quadrature.
The curves are the predictions from a QCD-inspired leading
order fit to the CCFR differential cross
section data with $R$= $R_{world}$ (for neutrino scattering) 
which does not include the HERMES effect. Large anomalous deviations 
from the fit (e.g. 200 - 300\%) are not seen in the CCFR data.

More details on this work can be found in reference 6 and 7.

\end{document}